\documentclass[showpacs,preprintnumbers,amsmath,amssymb,superscriptaddress,twocolumn,nofootinbib]{revtex4}
\usepackage{graphicx}
\usepackage{aas_macros}
\input{colordvi.tex}

\begin{document}
\title{Relationship between the CMB, SZ Cluster
Counts, and Local Hubble Parameter Measurements in a Simple Void Model}
\author{Kiyotomo Ichiki}
\email{ichiki@a.phys.nagoya-u.ac.jp}
\affiliation{%
Kobayashi-Maskawa Institute for the Origin of Particles and
the Universe, Nagoya University, Chikusa-ku, Nagoya, 464-8602, Japan
}
\affiliation{%
Department of Physics and Astrophysics, Graduate School of Science, Nagoya University, Nagoya
464-8602, Japan
}
\author{Chul-Moon Yoo}
\affiliation{%
Department of Physics and Astrophysics, Graduate School of Science, Nagoya University, Nagoya
464-8602, Japan
}
\author{Masamune Oguri}
\affiliation{%
Research Center for the Early Universe, the University of Tokyo, Tokyo 113-0033, Japan
}
\affiliation{%
Department of Physics, the University of Tokyo, Tokyo 113-0033, Japan
}
\affiliation{%
Kavli Institute for the Physics and Mathematics of the Universe (Kavli
IPMU, WPI), the University of Tokyo, Chiba 277-8583, Japan
}

\date{\today} \preprint{}
\begin{abstract}
 The discrepancy between the amplitudes of matter fluctuations inferred
 from Sunyaev-Zel'dovich (SZ) cluster number counts, the primary
 temperature, and the polarization anisotropies of the cosmic
 microwave background (CMB) measured by the Planck satellite can be
 reconciled if the local universe is embedded in an under-dense region
 as shown by Lee, 2014. Here using a simple void model assuming the
 open Friedmann-Robertson-Walker geometry and a Markov Chain Monte
 Carlo technique, we investigate how deep the local under-dense region
 needs to be to resolve this discrepancy. Such local void, if exists,
 predicts the local Hubble parameter value that is different from the
 global Hubble constant. We derive the posterior distribution of the
 local Hubble parameter from a joint fitting of the Planck CMB data
 and SZ cluster number counts assuming the simple void model. We show
 that the predicted local Hubble parameter value of $H_{\rm
   loc}=70.1\pm0.34~{\rm km\,s^{-1}Mpc^{-1}}$ is in better agreement
 with direct local Hubble parameter measurements, indicating that the local
 void model {may} provide a consistent solution to the
 cluster number 
 counts and Hubble parameter discrepancies.
\end{abstract}
\maketitle

\section{introduction}
The standard model of cosmology, also referred to the $\Lambda$CDM
model, is now well established based on a number of cosmological
observations. The simplest $\Lambda$CDM model contains six free
cosmological parameters, each of which has been determined with an
accuracy of a few percent through the measurement of Cosmic Microwave
Background (CMB) anisotropies by the Planck satellite
\cite{2014A&A...571A..16P}. The CMB anisotropies mainly manifest the
nature of density fluctuations in the universe at the recombination
era at $z\approx 1100$. It is therefore of great importance to
confront the standard model with observations of the local universe
($z\approx 0$) to confirm that the standard model is correct
throughout the history of the universe {(see section 5.6 of
\citep{2015arXiv150201589P} and references therein).}
%\cite{2015ApJ...799..214B}. 

Some authors, however, have reported cosmological measurements in the
local universe that conflict with the CMB result. For example, it has
been claimed that the amplitude of density fluctuations measured in
the local universe is systematically low compared with the prediction
from Planck's CMB data (e.g.,
\cite{2013PhRvL.111p1301M,2014PhRvL.112e1303B,2014PhRvL.112e1302W,2015PhRvD..91j3508B,2014PhRvL.113r1301S,2015MNRAS.451.2877M,2015PhRvD..92f3517L}). 
In particular, the number of massive clusters derived from the
Sunyaev-Zel'dovich (SZ) effect by Planck is about half that expected
from the CMB anisotropies \cite{2014A&A...571A..20P}. 
{This SZ result from Planck is consistent with the SZ-selected
cluster number counts from the Atacama Cosmology Telescope \cite{2013JCAP...07..008H} and the South Pole Telescope surveys \cite{2013ApJ...763..127R}.}
Perhaps related
to this discrepancy, the local Hubble parameter measured through the
traditional distance ladder tends to show a larger value (e.g., $H_0=
73.8 \pm 2.4~{\rm km\,s^{-1}Mpc^{-1}}$ by \cite{2011ApJ...730..119R})
than that inferred from the Planck result ($H_0\approx 67.3 \pm 1.0~{\rm 
km\,s^{-1}Mpc^{-1}}$). There have been many proposals to resolve these
discrepancies, including beyond-the-standard
models, e.g., massive neutrinos
\cite{2014PhRvL.112e1302W,2014PhRvL.112e1303B}, and decaying dark matter
\cite{2015arXiv150503644B}, and re-considerations of astrophysics and
calibration issues, e.g., the halo mass function
\cite{2015arXiv150207357B,2014MNRAS.440.2290M,2014MNRAS.439.2485C}, and mass
calibration
\cite{2015arXiv150908930B,2014MNRAS.443.1973V,2014A&A...570L..10A,2015ApJ...804..129Z,2015A&A...575A..30S}.

Here, we reinvestigate the idea of Lee 2014 \cite{2014MNRAS.440..119L},
where it was shown that the discrepancy can be resolved if we reside
in a local under-dense region. The idea is particularly interesting 
because the local under-dense region may explain the observed
discrepancy in the Hubble parameter as well \cite{2013PhRvL.110x1305M}.
In our analysis, we employ a simple Friedmann-Robertson-Walker (FRW)
model with an open geometry for the under-dense region and investigated
how deep the under-dense region is in order to resolve the discrepancy
using the Markov Chain Monte Carlo (MCMC) technique. We show that the
joint fitting of the CMB and SZ data set {\it predicts} the local 
Hubble parameter value. The posterior distribution of the local Hubble
parameter from the MCMC analysis is then confronted with direct local
Hubble parameter measurements to check the consistency of the local
void picture. 

The paper is organized as follows. In the next section we describe our
simple void model and our method of analysis. We present our
results and discussions in Section III. The final section is devoted
to our conclusion. The normalized Hubble parameter $h$ is defined by
$h\equiv H_0/(100~{\rm km\,s^{-1}Mpc^{-1}})$.

\section{Method}
\subsection{Local void model}

 Suppose that we live in a local (under-)dense region in a 
 background flat
 Friedmann-Robertson-Walker (FRW) universe. If the local region
 has a constant density profile, i.e., a top-hat profile, the
 evolution of the region is approximated by a slightly
 different FRW universe. 
 The local universe and the background universe 
 may be characterized by cosmological parameters:$(H_{\rm loc}, \Omega_{\rm loc})$ and $(H_{\rm bg}, \Omega_{\rm bg})$, respectively. 
 The local cosmological parameters $(H_{\rm loc}, \Omega_{\rm loc})$ 
 can differ in principle from those in the global FRW universe. Here,
 for simplicity,  
 let us assume that abundance ratios of the energy components in the
 universe are same in the global and the local under-dense regions by
 demanding that 
\begin{equation}
 \Omega_{\alpha, {\rm loc}}h_{\rm loc}^2 = \Omega_{\alpha}h_{\rm bg}^2~,
 \label{assum1}
\end{equation}
  where the subscript $\alpha=(\rm{v,c,b,\gamma,n})$ stands for the vacuum
 energy (v), CDM density (c), baryon density (b), photon density
 ($\gamma$) and neutrino density
 (n). 
 Note that we have assumed that the curvature parameter in the background 
 FRW model is equal to zero, i.e., $\Omega_{\rm K, bg}=0$. 
 The local curvature parameter, $\Omega_{\rm K, loc}$ is
 determined by
\begin{equation}
 \Omega_{\rm K, loc} = 1-\sum_i\Omega_{i, \rm{loc}}~,
\label{eq:curvature}
\end{equation}
so that the difference between $H_{\rm loc}$ and $H_{\rm bg}$ is compensated. 
{Then, we find $\Omega_{\rm K,loc}>0$ $(<0)$ for $H_{\rm loc}>H_{\rm
bg}$ $(H_{\rm loc}<H_{\rm bg})$. }
It should be noted that the present time must be fixed so that 
the present CMB temperature is 2.725K. 
Therefore, our assumption \eqref{assum1} implies that, 
for the present uniform density time slice, 
the CMB temperature we observe and that in the background universe 
are equally given by 2.725K. Equation \eqref{assum1} also implies that
the local void considered here is in the growing mode (see e.g., Eq.(34)
in \cite{2014PhRvD..89h3519L}).

Let us introduce a parameter $f_{\rm loc}$ as follows:
\begin{equation}
 H_{\rm loc} = f_{\rm loc}H_{\rm bg}~. 
\label{eq:fvoid}
\end{equation}
We are particularly interested in the case $f_{\rm loc}>1$ 
to explain the discrepancy in the measured values of 
the Hubble parameter. 
In this case, in order to change the present time slice from 
the uniform density slice 
to the uniform Hubble slice given by $H=H_{\rm loc}>H_{\rm bg}$, 
we have to go back in time and take the past slice in the background universe. 
Consequently, in the background universe, the density on this time slice 
is larger than that on the uniform density time slice. 
This means that the density profile has a void structure in the uniform Hubble slice. 
This observation is consistent with the result reported in 
Ref.~\cite{2013PhRvL.110x1305M}, which states that 
the local under-dense region may lead to a larger value of 
the local Hubble parameter. 
\footnote{
The correlation between the larger value of the local Hubble parameter and 
the local under-dense region can be 
also understood by considering conventional perturbation theory. 
 Usually, the initial condition for the Hubble flow of the top-hat
 region is set assuming that the density perturbation grows
 following linearized perturbation theory; in the radiation
 dominated era, the density contrast of region $\delta$ grows as
 $\delta \propto a^2$ in the synchronous gauge, where $a$ is the cosmic
 scale factor. In this case, the relationship between 
 the Hubble parameters of the top-hat region $H_{\rm loc}$ and the
 background universe $H_{\rm bg}$ is given by \cite{2012PhRvD..85f3521I}
%\begin{equation}
% H_{{\rm void},i} = H_{{\rm bg},i}
%  -\frac{2}{3}\frac{\delta_i}{1+\delta_i}H_{{\rm bg},i}~,
%\label{eq:eq1}
%\end{equation}
$$
 H_{{\rm loc},i} = H_{{\rm bg},i}
  -\frac{2}{3}\delta_i H_{{\rm bg},i}~,
$$
 where the
 subscript $i$ denotes a specific initial time. 
If the top-hat region is
 under-dense, i.e., $-1< \delta<0$, then $H_{{\rm loc},i}>H_{{\rm
 bg},i}$ and the inequality $H_{{\rm loc}}(t)>H_{{\rm
 bg}}(t)$ always holds true.
}

% We take a different approach because we mainly consider the local
% universe. Instead of setting the relationship between the under-dense
% region and the surrounding background universe in early times, we set
% this relation in the present by introducing a new parameter $f_{\rm
%   void}$ that relates the two Hubble parameters as
%%
%\begin{equation}
% H_{\rm loc} = f_{\rm loc}H_{\rm bg}~,
%\label{eq:fvoid}
%\end{equation}
%%
% where $H_{\rm bg}$ and $H_{\rm loc}$ are the Hubble parameters for the
% global and the local under-dense regions, respectively.  
% {\bf \Red{In the local open FRW universe, the cosmological parameters
% can differ in principle from those in the global FRW universe. Here,
% for simplicity,  
% let us assume that abundance ratios of the energy components in the
% universe are same in the global and the local under-dense regions by
% demanding that}}
% In the local,
% open FRW universe, the cosmological parameters are related to those in
% the global FRW model using the normalized Hubble parameter $h$ as
%
%\begin{equation}
% \Omega_{\alpha, {\rm loc}}h_{\rm loc}^2 = \Omega_{\alpha}h_{\rm bg}^2~,
%\end{equation}
%%
% where the subscript $\alpha=(\rm{v,c,b,n})$ stands for the vacuum
% energy (v), CDM density (c), baryon density (b), and neutrino density
% (n). The local curvature parameter, $\Omega_{\rm K, void}$ is
% determined by
%
%\begin{equation}
% \Omega_{\rm K, void} = 1-\sum_i\Omega_{i, \rm{loc}}~.
%\label{eq:curvature}
%\end{equation}
%

In this paper, we calculate the cluster abundance 
based on the local cosmological parameters: $(H_{\rm loc}, \Omega_{\rm loc})$, 
which are fixed by $(H_{\rm bg}, \Omega_{\rm bg})$ and $f_{\rm loc}$
through Eqs.~\eqref{assum1} and \eqref{eq:fvoid}. 
{In other words, we assume that the local void
covers up the entire region where the Planck clusters were 
found, namely, $z\lesssim 1$, or in terms of the radial comoving distance, $\lesssim
3000$ Mpc.} 

\begin{figure}[h]
\centering
\includegraphics[width=0.5\textwidth]{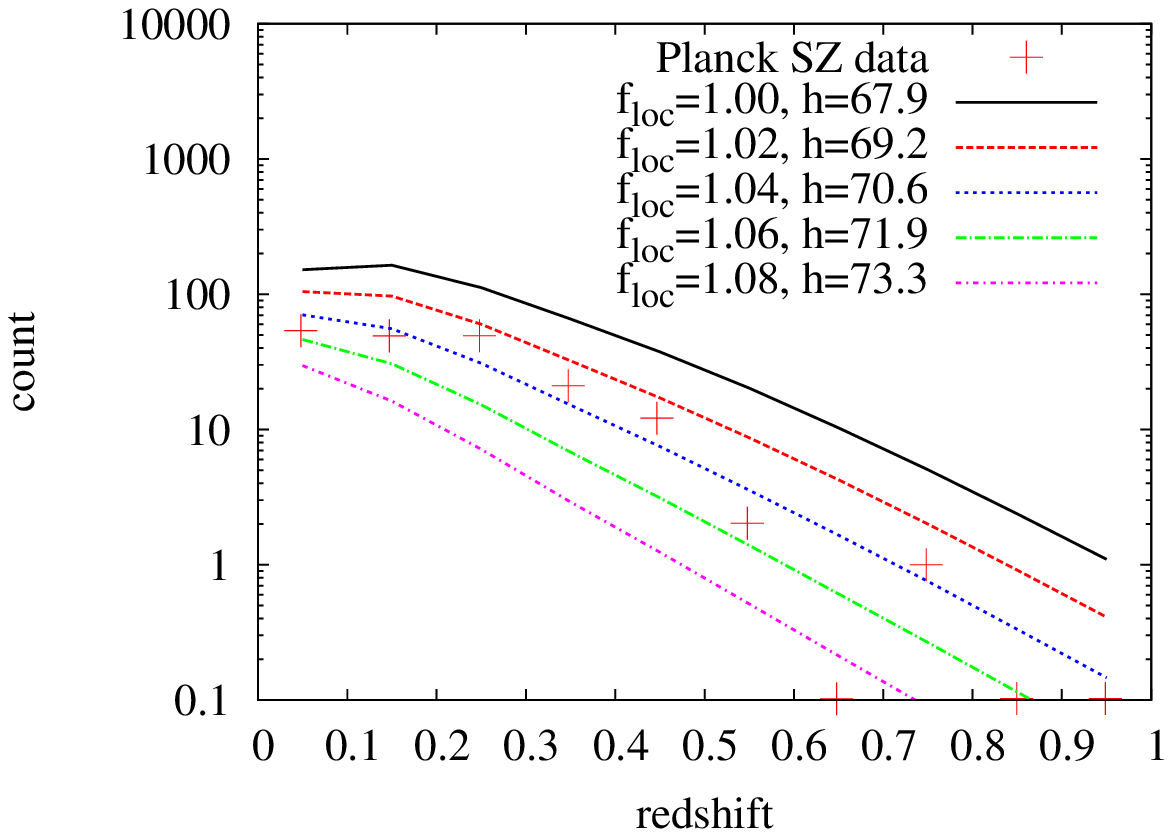}
\caption{SZ cluster number counts predicted in the local void model
  are compared with the observed number counts from the Planck 2013
  result \cite{2014A&A...571A..20P}. The parameter $f=f_{\rm loc}$ defined
  in Eq.~(\ref{eq:fvoid}) describes the difference between the local
  and global Hubble parameters. The parameter $h$ here denotes the
  normalized local Hubble parameter value corresponding to each
  $f_{\rm loc}$.}  \label{fig:fig1} 
\end{figure}

The effect of the parameter $f_{\rm loc}$ on the cluster abundance is
shown in Fig.~\ref{fig:fig1}.  The decrease in the cluster abundance
with increasing $f_{\rm loc}$ mainly comes from a smaller amplitude of
density fluctuations at corresponding scales due to a slower growth rate
in the open FRW model than that in the flat FRW one. Following the
convention, we denote the fluctuation amplitude by $\sigma_{8,
\rm{loc}}$, the root-mean-squared of linear fluctuations within a
top-hat sphere of $8\,h^{-1}$~Mpc radius. Assuming that primordial
density fluctuations have the same initial power spectrum both in the
under-dense region and the surrounding background universe, the
parameter $\sigma_{8, \rm{loc}}$ is derived from the other cosmological
parameters.

For simplicity, we assume that the angular diameter distance to the
last scattering surface remains unchanged. This can be justified if the
density profile in the high redshift universe, say in $2 \lesssim z
\lesssim 1000$, is modified to compensate for the stretch in distance
by the local under-dense region. In this case, the Hubble parameter
$H_{\rm bg}$ is not directly observed but inferred from the CMB
measurements. On the other hand, $H_{\rm loc}$ can directly be
observed using local distance measurements. 

\begin{table}[htbp]
\begin{tabular}{cc}
\hline
\hline
{$z$}  &  {$M_{\rm min}(z)$ [$M_\odot$]} \\
\hline
 0.05 & 3.12e14 \\
 0.15 & 6.15e14 \\
 0.25 & 8.20e14 \\
 0.35 & 9.66e14 \\
 0.45 & 1.06e15 \\
 0.55 & 1.13e15 \\
 0.65 & 1.17e15 \\
 0.75 & 1.20e15 \\
 0.85 & 1.21e15 \\
 0.95 & 1.21e15 \\
\hline
\hline
\end{tabular}
\caption{Mean minimum masses of the Planck 2013 SZ cluster sample as a
  function of redshift, which are taken from
  \cite{2014A&A...571A..20P}. {These masses are defined at 50\%
 completeness on average for the unmasked sky.}}
\label{tb:minimum_mass}
\end{table}

\subsection{Mass function and cluster abundance}
To predict the abundance of SZ clusters, we adopt the fitting form of
the mass function $dn_{\rm halo}/dM$ presented in \cite{2008ApJ...688..709T}. 
We adopt the parameters $D_{\rm crit}=500$ and $D_{\rm cluster}=D_{\rm
  crit}/\Omega_{\rm M}(z)$, where $\Omega_{\rm M}(z)$ is the total
matter density over the critical density at redshift $z$.  The cluster
abundance of mass $M$ at redshift $z$ is given by
\begin{equation}
 \frac{d N}{dz} (z)= f_{\rm sky}\int_0^\infty dM
 \chi(M)
 \frac{dN}{dM}(M,z)\frac{dV(z)}{dz}~, 
\end{equation}
where the co-moving volume $V(z)$ is given by 
\begin{equation}
 \frac{dV(z)}{dz}=4\pi (1+z)^2 \frac{d_A(z)^2}{H(z)}~,
\end{equation}
and the sky fraction that is covered by the Planck SZ cluster counts is
fixed to $f_{\rm sky}=0.65$. The function $\chi(M)$ represents the survey
completeness as a function of halo mass $M$ that takes account of the
fact that some fraction of clusters falls out from detection due to
the survey strategy, detection limit, and so on.
We simply write this function as \cite{2013JCAP...12..012C} 
\begin{equation}
 \chi(M) = \int_{M_{\rm min}(z)}^\infty dM^\prime P(M^\prime|M)~,
\end{equation}
where the function $P(M^\prime|M)$ describes the probability that the
mass is estimated as $M^\prime$ for a cluster with a true mass $M$,
which for example originates from the scatter in the scaling relation
between halo masses and SZ signals. In this analysis, we assumed
that it follows a log-normal distribution with variance 
$\sigma^2_{\ln M}$ as 
\begin{equation}
 P(M^\prime|M) = \frac{M^{\prime -1}}{\sqrt{2\pi \sigma^2_{\ln M}}}\exp
\left[ - \frac{(\ln M^\prime -\ln M)^2}{2\sigma^2_{\ln M}}\right]~.
\end{equation}
We fix $\sigma_{\ln M}=0.2$, which roughly reproduces the theoretical
curve for the number counts of Planck SZ clusters presented in
\cite{2014A&A...571A..20P}, as shown by the black curve in 
Fig.~\ref{fig:fig1}.  The minimum masses obtained from the Planck SZ
survey are listed in Table~\ref{tb:minimum_mass}.

{A caveat is that, while we have used the Tinker
  mass function to compare our results with those of the Planck
  fiducial analysis, the fitting function is not very well tested in
  an open $\Lambda$CDM model which is of our main interest in this
  paper. We expect that the fitting function works fine even in
  non-flat universes as long as important ingredients of the fitting
  function such as the matter density, linear power spectrum, and
  linear growth rate are properly included, but this assumption should
  be carefully validated by e.g., $N$-body simulation.}

\subsection{MCMC}
 In our MCMC analysis we use both the Planck CMB data and the SZ cluster
 number counts. Specifically, we used a flat $\Lambda$CDM model
 when fitting to the CMB data with the six standard parameters, while we
 used an open $\Lambda$CDM model with an extra parameter $f_{\rm loc}$
 to fit to the SZ cluster data;
 the other parameters were derived from the relations in
 Eqs.~(\ref{eq:fvoid})--(\ref{eq:curvature}). We use a modified
 version of CosmoMC \cite{2002PhRvD..66j3511L} to explore the
 likelihoods. For the likelihood of the cluster number counts, we
 assume that the number of the cluster obeys Poisson statistics
 \cite{1979ApJ...228..939C}, which is a good approximation for massive
 clusters as considered here \cite{2003ApJ...584..702H}.

\section{Results and Discussion}
\begin{figure}[h]
\centering
\includegraphics[width=0.5\textwidth]{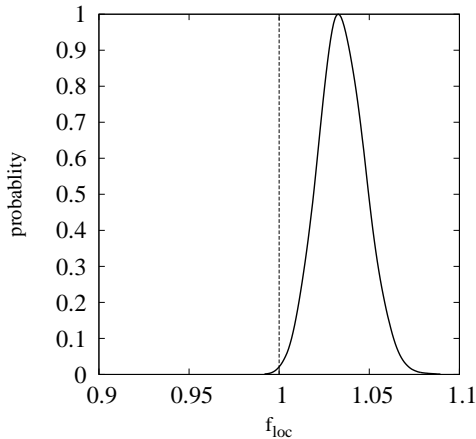}
\caption{One-dimensional posterior probability distribution of the
  parameter $f_{\rm loc}$ (see Eq.~\ref{eq:fvoid}) from the joint
  fitting of the CMB and SZ cluster number counts. The dotted line
  ($f_{\rm loc}=1$) denotes the standard cosmological model without
  any local over- or under-dense region.}
\label{fig:fig2}
\end{figure}

\begin{figure}[ht]
\centering
\begin{minipage}[m]{0.23\textwidth}
\includegraphics[width=1.05\textwidth]{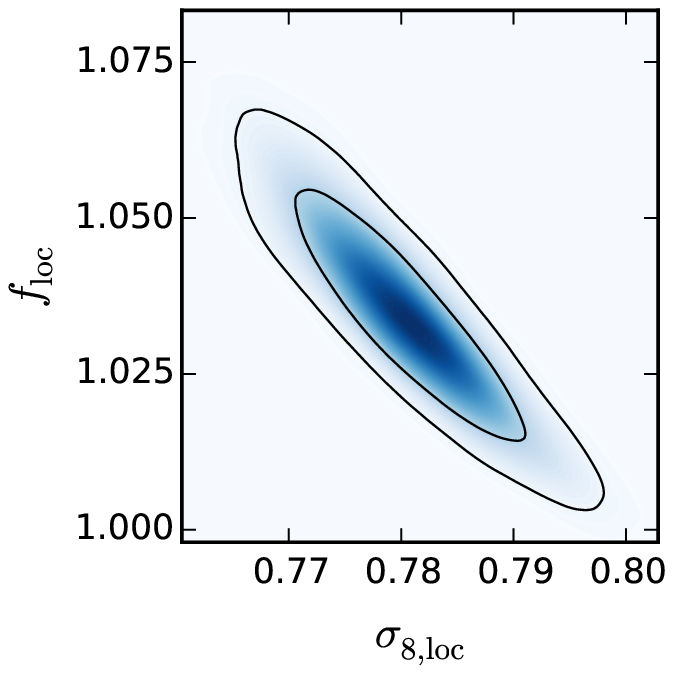} 
\end{minipage}
\begin{minipage}[m]{0.23\textwidth}
\includegraphics[width=1.05\textwidth]{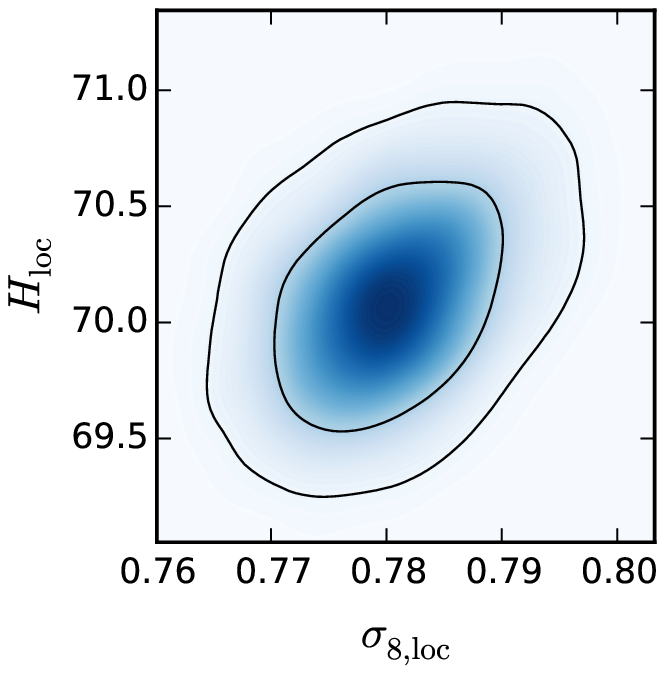} 
\end{minipage}
\caption{Posterior probability distribution on the $f_{\rm
    loc}-\sigma_{8,{\rm loc}}$ ({\it left}) and $H_{\rm
    loc}-\sigma_{8,{\rm loc}}$ ({\it right}) planes.}
\label{fig:fig2-2}
\end{figure}

 First, we show the posterior distribution for $f_{\rm loc}$ from
 the joint fitting of the CMB and SZ number counts in
 Fig.~\ref{fig:fig2}. As shown in the Figure, the parameter $f_{\rm
   loc}$ tends to be $f_{\rm loc}>1$ indicating that the model with
 a local under-dense region fits better with the data set. The result
 $f_{\rm loc}=1.03 \pm 0.01245$ indicates that the standard
 $\Lambda$CDM model is excluded at $2.4 \sigma$ level. 
 We find the improvement of $\chi^2$ to be $\Delta \chi^2
 \approx 8.44$ with one additional parameter, which is, in terms of
 Akaike's 
 information criterion, ${\rm AIC}\approx 6.44$. This indicates that the
 void model is preferable to the $\Lambda$CDM model for the combined
 CMB and SZ data set. The two
 dimensional constraint on the $f_{\rm loc}-\sigma_{8,{\rm loc}}$
 plane is shown in the left panel of Fig.~\ref{fig:fig2-2}. It is
 evident that the trend for $f_{\rm loc}>1$ is required in order to
 realize the smaller $\sigma_8$ as inferred from the measurements of
 the local universe.   

An interesting relation is found in the right panel of
Fig~\ref{fig:fig2-2}, which shows a positive correlation between
$\sigma_{8,{\rm loc}}$ and $H_{\rm loc}$. This correlation is
non-trivial because a larger $f_{\rm loc}$ leads a smaller
$\sigma_{8,{\rm loc}}$ but a larger $H_{\rm loc}$ if the other
cosmological parameters are fixed. We find that this correlation is
attributed to the well-known $\sigma_8$--$\Omega_m$ degeneracy in the
cluster abundance. In order to produce the same cluster abundance, a
larger $\sigma_{8,{\rm loc}}$ requires a smaller $\Omega_{m,
  \rm{loc}}$, hence a larger $H_{\rm loc}$ because  
$\Omega_{m, \rm{loc}} h_{\rm loc}^2 = \Omega_m h^2_{\rm bg}$ is
tightly constrained from the CMB.

\begin{figure}[h]
\centering
\begin{minipage}[m]{0.23\textwidth}
\includegraphics[width=1.15\textwidth]{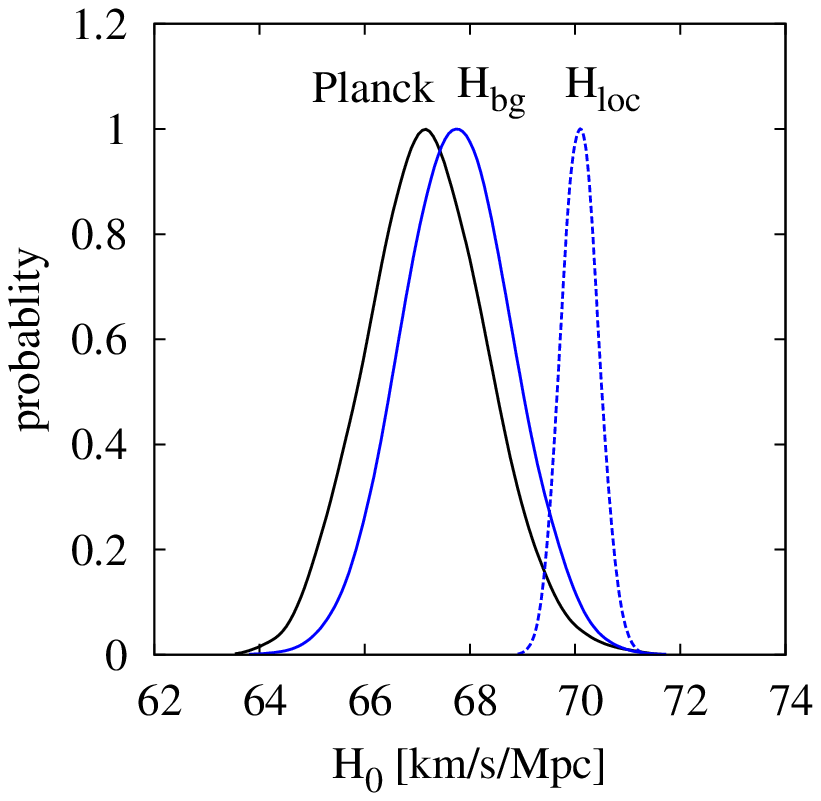} 
\end{minipage}
\begin{minipage}[m]{0.23\textwidth}
\includegraphics[width=1.15\textwidth]{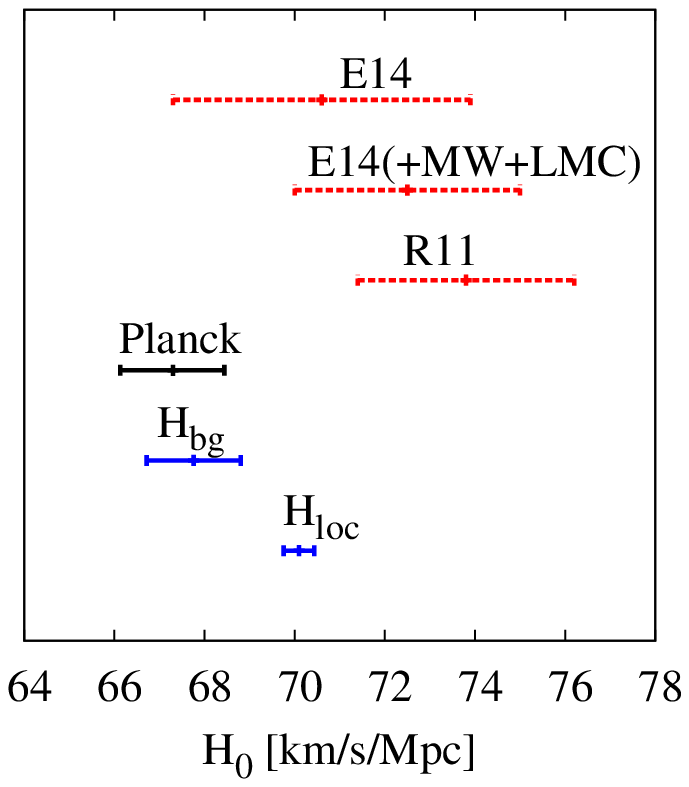} 
\end{minipage}
\caption{
 ({\it Left}) Probability distribution functions for the Hubble parameter
 from Planck (black solid line), and $H_{\rm bg}$ (blue solid) and
 $H_{\rm loc}$ (blue dashed) from the combined CMB
 and SZ data set.
 ({\it Right}) Constraints on the Hubble parameter are compared with the direct
   local estimates for the Hubble constant by Riess et al. (R11), and by
 Efstathiou using NGC 4258 (E14) and plus other distance anchors
 (E14+MW+LMC). The black point with an error bar shows the constraint
 from Planck CMB. In
 the local void model, Hubble parameters are separated into local
 ($H_{\rm loc}$) and global ($H_{\rm bg}$) parameters, as shown in the
 figure. 
 % The separation of the Hubble parameters between
 %  Planck (black solid-line) and the Planck and SZ combination (red
 %  dashed-line) indicates the tension between these two data sets. In
 %  the local void model, Hubble parameters are separated into local
 %  ($H_{\rm local}$; blue dashed) and global ($H_{\rm CMB}$; blue
 %  solid) parameters.
} 
\label{fig:fig3}
\end{figure}

 A key feature of this local void model is that the local Hubble
 parameter defined in Eq.~(\ref{eq:fvoid}) is automatically adjusted
 as a result of fitting to the observed SZ cluster number counts. 
Figure~\ref{fig:fig3} shows the posterior probability distributions of
the Hubble parameters with and without the void parameter $f_{\rm
  loc}$, together with the local Hubble parameter measurements
  from Riess et al. \cite{2011ApJ...730..119R} and Efstathiou \cite{2014MNRAS.440.1138E}. The main results are summarized as follows. By keeping the
void parameter at the standard $\Lambda$CDM value ($f_{\rm loc}=1$),
the Planck CMB gives $H_0=67.3 \pm 1.2~{\rm km\,s^{-1}Mpc^{-1}}$
whilst the CMB and SZ combination gives $H_0=70.7 \pm 0.3~{\rm
  km\,s^{-1}Mpc^{-1}}$, indicating the disparity between these two
data sets. We find that in the current void model, the local Hubble
parameter is predicted by the joint fitting of the CMB and SZ cluster
number counts as $H_{\rm loc}=100 h_{\rm loc}=70.1 \pm 0.34~{\rm km\,s^{-1}Mpc^{-1}}$.  
This value is in better agreement with the local Hubble measurements
by Riess et al. \cite{2011ApJ...730..119R}, and interestingly, is in
excellent agreement with the value $H_0=70.6\pm 3.3~{\rm
  km\,s^{-1}Mpc^{-1}}$ recently derived by Efstathiou using the
updated geometric maser distance to NGC4258 
\cite{2014MNRAS.440.1138E}. 

There are several other possibilities to resolve the discrepancy. For
example it may be possible that the mass bias factor is significantly
lower than the inferred value, i.e., masses of clusters observed by
Planck SZ is in fact larger than those inferred by the scaling relation,
and therefore, they are less abundant in the universe
\cite{2014MNRAS.443.1973V}. Since the 
publishing of their results in 2013, the Planck collaboration has
updated their bias calibration using cosmic shear and CMB lensing
measurements, and their {2015 estimates give
similar mass bias 
values to that in the baseline model assumed in 2013
\cite{2015arXiv150201597P}. The results of cluster number counts in 2013
and 2015 are consistent with each other if
the same mass bias is assumed.}  Independent analysis also shows that the
mass bias cannot fully resolve the discrepancy
\cite{2015A&A...579A...7R,2015arXiv150704493O,2015MNRAS.449..685H}.
{Recently, an important update has been made in
Battaglia et al. \cite{2015arXiv150908930B}. They found that an Eddington bias correction, if it is applied to
the weak lensing mass calibration analysis of the Planck SZ clusters,
brings the mass bias factors measured in von der Linden et
al. \cite{2014MNRAS.443.1973V} and Hoekstra et
al. \cite{2015MNRAS.449..685H} closer to the values that are
small enough to explain both the Planck CMB and SZ results. Further
analysis of this systematic uncertainty is of primary importance for
understanding the origin of the discrepancy.}

{From a cosmological point of view,} while massive
neutrinos may be an interesting possibility
\cite{2014PhRvL.112e1302W,2014PhRvL.112e1303B}, a larger neutrino mass
generally demands a smaller Hubble parameter, which makes the
discrepancy between the global and local Hubble parameters even worse
 \cite{2010JCAP...03..015S}. Other possibilities include modification
 of the mass function due to baryonic effects
\cite{2015arXiv150207357B,2014MNRAS.440.2290M,2014MNRAS.439.2485C},
and decaying dark matter \cite{2015arXiv150503644B}.  Non-Gaussianity
may also explain the discrepancy, although the Planck bi-spectrum already
put severe constraints \cite{2015arXiv150201592P}, making this
explanation difficult as long as the non-Gaussianity is
scale-independent \cite{2013MNRAS.435..782T}.

Our current study needs to be advanced in several directions. The large
bulk flow associated with the local void generates secondary CMB
anisotropies {through the kinetic Sunyaev-Zeldovich effect (kSZ).
Here we give a rough estimate of the kSZ effect arising from the local
void following Moss et al. \cite{2011PhRvD..83j3515M}.  Suppose that the dipole induced by the
radial bulk velocity is the dominant anisotropy seen by the scatterer
(i.e., clusters). We estimate the amplitude of this dipole from the
difference in redshift between
incoming and outgoing CMB photons felt for the same period in conformal time.  The dipole anisotropy at clusters is then given by
\begin{equation}
\frac{\Delta T}{T} \approx  1 - \frac{R(\eta_{*})}{a(\eta_{*})}~,
\end{equation}
where $a(\eta_\ast)$ and $R(\eta_\ast)$ denote the scale factors of the
background and void regions when the incoming photon enters into the
void, and follow the Friedmann equations in flat and open $\Lambda$CDM
models, respectively. For our void model that gives $H_{\rm loc}=70$, we
obtain $\Delta T/T\approx 7\times 10^{-4}$ for $a(\eta_\ast)\approx
0.01$ and $\Delta T/T\approx 9\times 10^{-3}$ for $a(\eta_\ast) \approx
0.1$. Therefore, a fairly large void seems necessary to avoid the
constraint from peculiar velocities measured by the Planck
collaboration, which gives $\Delta T/T \lesssim 6.4\times 10^{-4}$
(2$\sigma$) \cite{2014A&A...561A..97P}. These numbers above are order of magnitude
estimates; further detailed investigation would be
needed for a more quantitative constraint by using concrete models of
spherically symmetric inhomogeneous universes, e.g., generalized
$\Lambda$-Lema\^itre-Tolman-Bondi dust models \cite{1997GReGr..29..641L,1934PNAS...20..169T,1947MNRAS.107..410B}.} 
The effect on other
probes such as baryon acoustic oscillation and cosmic shear measurements
should also be investigated in order to see whether the local void is a
viable cosmological model.

This study is based on the Planck 2013 results. Interestingly, the discrepancy
persists in the Planck 2015 results \cite{2015arXiv150201597P} 
and therefore our results should still be as valid for the latest data set.

\section{Conclusion}
In this paper, we have considered a simple FRW model with an open
geometry to investigate the magnitude of the effect of the local
under-dense region on the number of SZ clusters. We have shown that a
local under-dense region may be responsible for the discrepancy
between cosmological parameters from SZ cluster number counts and the
CMB. If we attribute the discrepancy to the local void, the local
Hubble parameter is predicted as $H_0=70.1\pm 0.34~{\rm
  km\,s^{-1}Mpc^{-1}}$, which is 
in better agreement with direct local Hubble parameter measurements. Our result
indicates that the local void model {may serve} as a new concordance model
that explains the CMB, SZ cluster number counts, and local
Hubble parameter measurements simultaneously.

% \begin{table}[htbp]
% \begin{tabular}{cc}
% \hline
% \hline
% {parameter}  &  {values} \\
% \hline
%  $\Omega_b h^2$ & 0.022068 \\
%  $\Omega_c h^2$ & 0.12029 \\
%  $h$ & 67.854 \\
%  $\Omega_\nu h^2$ (1 massive nu)& 0.00064514 \\
%  $A_s$ & 2.21536e-9 \\
%  $n_s$ & 0.9624 \\
%  $\tau$ & 0.0925 \\
% \hline
% \hline
% \end{tabular}
% \caption{cosmological parameters} 
% \label{tb:params}
% \end{table}

\acknowledgments 
The authors would like to thank Matteo Costanzi for his kind correspondence.
Our thanks also go to the anonymous referee for giving us helpful
comments which improve our paper.
This work has been supported in part by Grant-in-Aid for
Scientific Research No. 24340048 and No. 26800093 from the Ministry of
Education, Sports, Science and Technology (MEXT), Japan.
This work was supported in part by World Premier International
Research Center Initiative (WPI Initiative), MEXT, Japan. 

%\appendix*
%\section{appendix}
%%%%%%%%%%%%%%%%%%%%%%%%%%%%%%%%%%%%%%%%%%%%%%%%%%%%%%%%%%%%%%%%%

\bibliography{paper}

\end{document}